\documentclass[review]{elsarticle}

\usepackage{amssymb}
\usepackage{amsmath}
\usepackage{url}
\usepackage{graphicx}%
\usepackage{multirow}%
\usepackage{amsmath,amssymb,amsfonts}%
\usepackage{amsthm}%
\usepackage{mathrsfs}%
\usepackage[title]{appendix}%
\usepackage{xcolor}%
\usepackage{textcomp}%
\usepackage{manyfoot}%
\usepackage{booktabs}%
\usepackage{algorithm}%
\usepackage{algorithmicx}%
\usepackage{algpseudocode}%
\usepackage{listings}%

\usepackage{lineno}

\journal{Materials Today}

\begin{document}

\begin{frontmatter}



\title{Epistemic Closure: Autonomous Mechanism Completion for Physically Consistent Simulation} 


\author[inst1]{Yue Wu\fnref{fn1}}

\author[inst1]{Tianhao Su\fnref{fn1}}

\author[inst1,inst3]{Rui Hu}

\author[inst4]{Mingchuan Zhao}

\author[inst1,inst2]{Shunbo Hu\corref{cor1}}
\ead{shunbohu@shu.edu.cn}

\author[inst1]{Deng Pan\corref{cor1}}
\ead{DPan\_MGI@shu.edu.cn}

\author[inst1,inst2]{Jizhong Huang\corref{cor1}}
\ead{hjizhong@shu.edu.cn}


\affiliation[inst1]{organization={Materials Genome Institute, Shanghai University},
            city={Shanghai},
            postcode={200444},
            country={China}}

\affiliation[inst2]{organization={Institute for the Conservation of Cultural Heritage, School of Cultural Heritage and Information Management, Shanghai University},
            city={Shanghai},
            postcode={200444},
            country={China}}

\affiliation[inst3]{organization={Shanghai Swave Technology Co., Ltd},
            city={Shanghai},
            country={China}}

\affiliation[inst4]{organization={School of Mathematical Sciences, University College Cork},
            city={Cork},
            country={Republic of Ireland}}

\cortext[cor1]{Corresponding author.}
\fntext[fn1]{These authors contributed equally to this work.}

\begin{abstract}
The integration of Large Language Models (LLMs) into scientific discovery is currently hindered by the \textbf{"Implicit Context"} problem, where governing equations extracted from literature contain invisible thermodynamic assumptions (e.g., undrained conditions) that standard generative models fail to recognize. This leads to \textbf{"Physical Hallucination"}: the generation of syntactically correct solvers that faithfully execute physically invalid laws. Here, we introduce a Neuro-Symbolic Generative Agent that functions as a cognitive supervisor atop traditional numerical engines. By encapsulating physical laws into modular \textbf{"Constitutive Skills"} and leveraging latent intrinsic priors, the Agent employs a Chain-of-Thought reasoning workflow to autonomously validate, prune, and complete physical mechanisms. We demonstrate this capability on the challenge of thermal pressurization in low-permeability sandstone. While a standard literature-retrieval baseline erroneously predicts catastrophic material failure by blindly adopting a rigid "undrained" simplification, our Agent autonomously identifies the system as operating in a drained regime (Deborah number $De \ll 1$) via dimensionless scaling analysis. Consequently, it inductively completes the missing dissipation mechanism (Darcy flow) required to satisfy boundary constraints, predicting a stable stress path consistent with experimental reality. This work establishes a paradigm where AI agents transcend the role of coding assistants to act as epistemic partners, capable of reasoning about—and correcting—the theoretical assumptions embedded in scientific data.
\end{abstract}

\begin{graphicalabstract}
    \centering
    \includegraphics[width=1.0\linewidth]{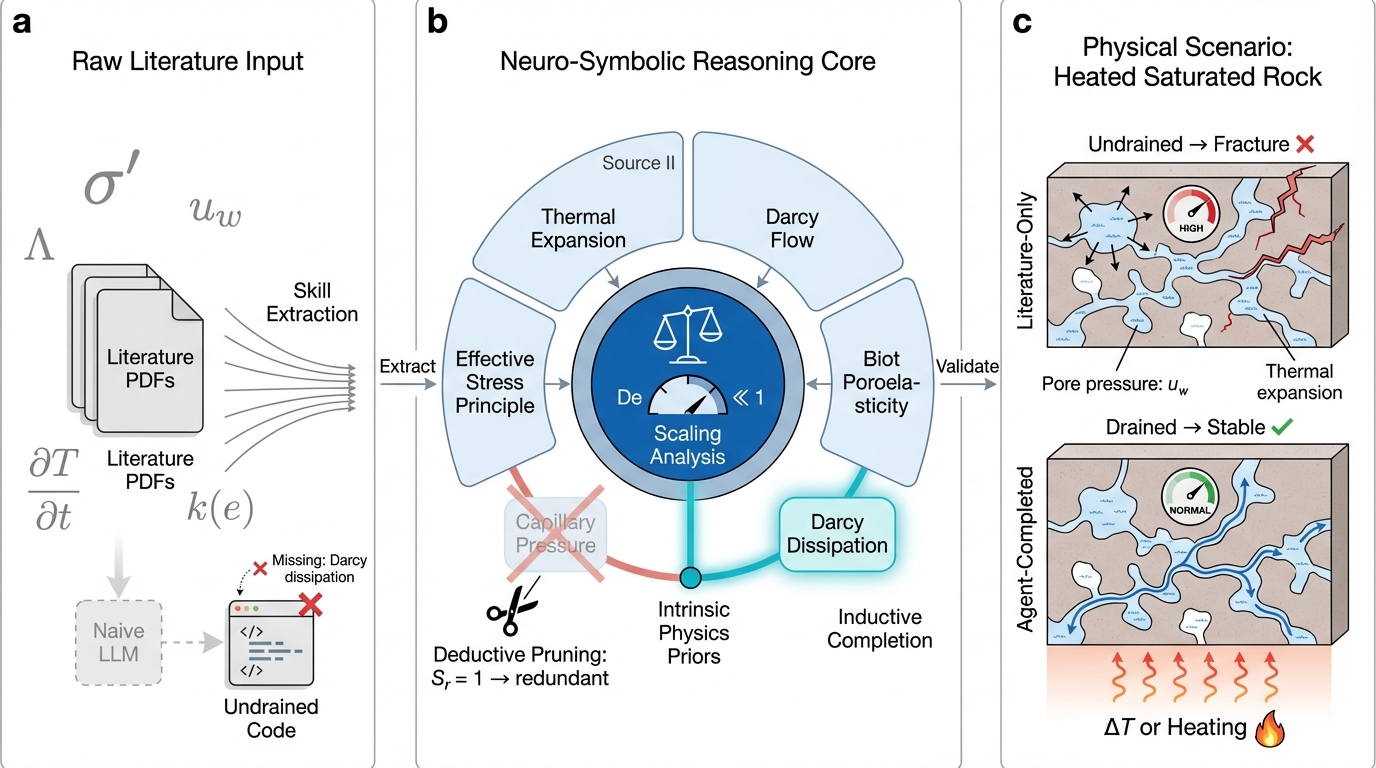} 
\end{graphicalabstract}


\begin{keyword}
Neuro-Symbolic AI \sep Physical Hallucination \sep Autonomous Mechanism Completion



\end{keyword}

\end{frontmatter}



\section{Introduction}

The integration of Artificial Intelligence into scientific discovery has transcended the phase of static property prediction (e.g., AlphaFold for protein structure \cite{jumper2021highly}, GNoME for material stability \cite{merchant2023scaling}) and is now confronting the dynamic frontier of multi-physics simulation\cite{raissi2019physics, li2021burigede, lu2021learning, wang2023scientific}. While traditional numerical solvers (FEM/FVM) demand rigorous, explicit definition of governing laws\cite{zienkiewicz2005finite, versteeg2007introduction}, the scientific knowledge required to formulate these laws remains fragmented across unstructured literature. The ultimate goal of "AI Scientist" agents is to bridge this epistemic gap: autonomously reading literature, synthesizing mathematical models, and executing high-fidelity simulations without human intervention\cite{boiko2023autonomous, m2024augmenting, wang2023voyager, park2023generative}.

The realization of this vision has been catalyzed by the recent performance leap in General-Purpose Large Language Models (LLMs), such as Claude 4.5 Sonnet\cite{anthropic2024claude} and Gemini 3 Pro\cite{team2023gemini}. Unlike their predecessors, these frontier models exhibit emergent reasoning capabilities that extend beyond syntactic translation to semantic understanding \cite{wei2022emergent, bubeck2023sparks, wei2022chain, yao2023tree}. Coupled with the rise of Agentic Frameworks—which structure LLM outputs into executable "Skills" or tools\cite{yao2022react, schick2023toolformer, xi2025rise, shen2023hugginggpt}—it is now computationally feasible to construct autonomous systems that can manipulate abstract physical concepts as modular programming units\cite{brohan2023can, maeureka}.

However, a critical bottleneck remains: the "Implicit Context" problem. Scientific equations in literature are rarely self-contained; they are bounded by invisible assumptions (e.g., "undrained conditions," "isothermal," or "small strain") that are obvious to a human expert but opaque to a standard "Copilot"\cite{ji2023survey}. A naive LLM, treating physics as a text completion task, often falls into the trap of "Physical Hallucination": generating code that is syntactically correct and mathematically solvable, but physically disastrous because it applies a valid law (e.g., undrained pressurization) outside its validity domain (e.g., a drained boundary)\cite{wang2023scibench, lewkowycz2022solving, taylor2022galactica}.

This limitation highlights a structural gap in the current computational landscape. Established multi-physics platforms like MOOSE \cite{gaston2009moose}, FEniCSx \cite{baratta_2025_18101307}, or JAX-FEM \cite{xue2023jax} have achieved high maturity in automating downstream numerical execution (e.g., weak-form assembly, linearization). However, they fundamentally operate as "agnostic calculators": they lack the cognitive agency to question the validity of their inputs. They rely entirely on an external entity—traditionally a human expert—to perform the upstream decision-making: selecting the correct constitutive laws, verifying thermodynamic consistency, and determining the appropriate coupling topology\cite{belytschko2014nonlinear}. Currently, this "Cognitive Supervisor" role remains manual; if the solver is fed a mathematically valid but physically incomplete model (as generated by a naive LLM), it will faithfully converge to a physically erroneous solution.

To bridge this gap, we introduce a Neuro-Symbolic Generative Agent\cite{udrescu2020ai, krenn2022scientific, sarker2022neuro} that elevates the LLM from a code translator to a Reasoning Kernel, effectively automating the role of the Cognitive Supervisor. Our framework treats extracted equations not as static text, but as "Constitutive Skills"—encapsulated logic units with semantic metadata. By leveraging the intrinsic physical priors embedded within the LLM's latent space (verified via knowledge graph extraction), the Agent executes a Chain-of-Thought (CoT) workflow to:
\begin{enumerate}
\item \textbf{Deductively Prune} redundant mechanisms (e.g., identifying capillary pressure as negligible in saturated states).
\item \textbf{Inductively Complete} missing physics (e.g., synthesizing Darcy flow dissipation when dimensionless scaling analysis reveals a drained regime).
\end{enumerate}

We validate this approach on a representative geomechanical challenge: thermal pressurization in low-permeability rock (Rothbach sandstone). This problem serves as an ideal "cognitive stress test" because it involves competing timescales of heat generation and fluid dissipation. We demonstrate that while a standard literature-retrieval baseline fails by blindly adopting an "undrained" simplification, our Agent autonomously identifies the correct physical regime, injects the necessary dissipation terms, and predicts a stable stress path consistent with experimental reality.

\section{Results}\label{sec3}

\begin{figure}[htbp]
    \centering
    \includegraphics[width=\textwidth]{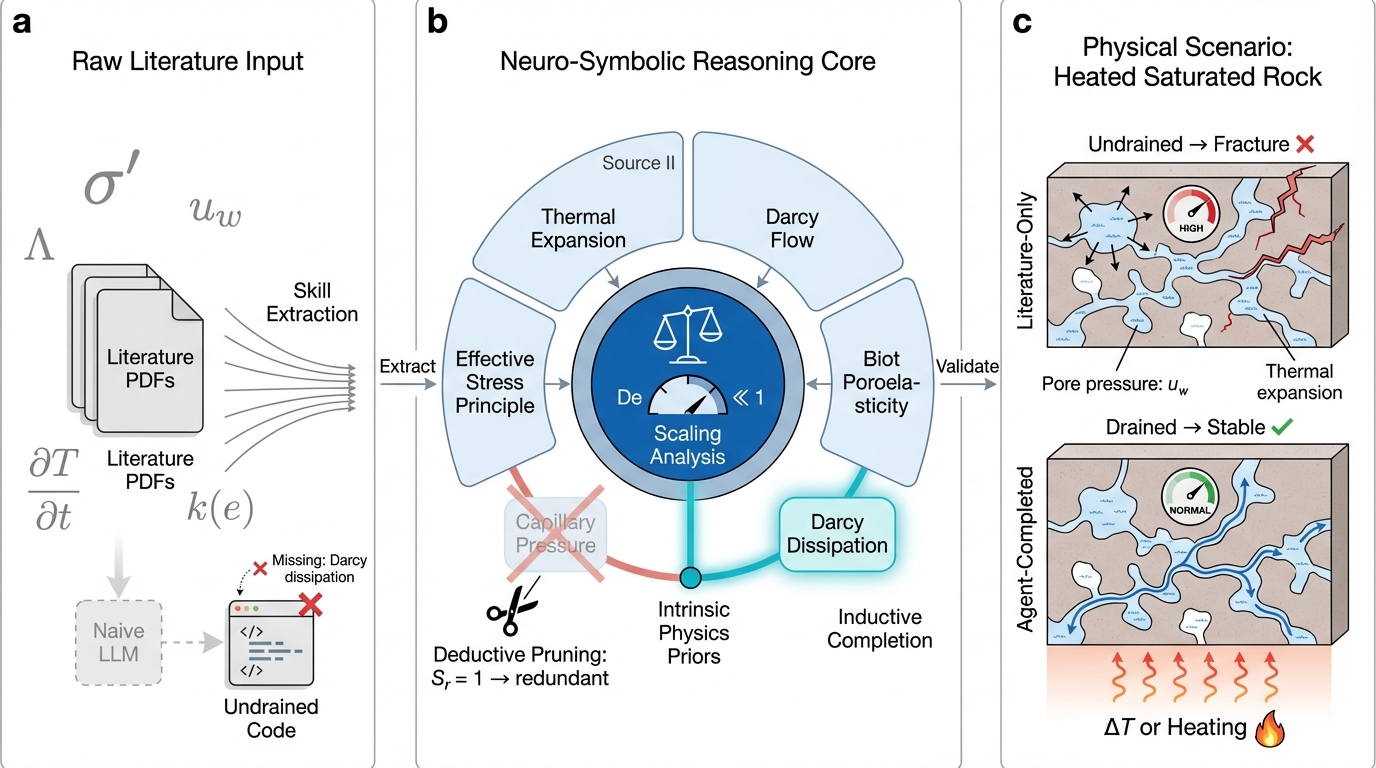}
    \caption{\textbf{Architectural overview of the Neuro-Symbolic Generative Agent.} 
    \textbf{a,} From Text to Skills: The Agent ingests unstructured scientific literature (PDFs), extracting governing equations into modular "Constitutive Skills." Unlike a naive LLM (grey path) that blindly executes retrieved code, our Agent treats these skills as candidate hypotheses.
    \textbf{b,} The Reasoning Core: The Agent acts as a cognitive gatekeeper. It employs \textbf{Deductive Pruning} (Scissors icon) to remove mathematically redundant mechanisms (e.g., capillary pressure in saturated states) and \textbf{Inductive Completion} (Blue path) to inject missing physics. The decision logic is governed by \textbf{Dimensionless Scaling Analysis} (Central Dial), specifically using the Deborah number ($De$) to arbitrate between competing mechanisms.
    \textbf{c,} Physical Consequence: A comparison of simulation outcomes for heated saturated rock. The "Literature-Only" model (top), constrained by the extracted "undrained" assumption, predicts false material fracture (Physical Hallucination). In contrast, the "Agent-Completed" model (bottom) correctly identifies the drained regime, activating Darcy flow to relieve pressure and predicting a stable, physically valid state.}
    \label{fig1}
\end{figure}

To validate the proposed framework, we deploy the Agent to solve a representative geomechanical challenge: thermal pressurization in low-permeability rock (Rothbach sandstone). This problem requires balancing thermodynamic driving forces with hydraulic dissipation, making it an ideal testbed for the Agent's reasoning capabilities. 
The implementation follows a three-stage autonomous workflow:
\begin{enumerate}
    \item \textbf{Skill Ingestion (Figure \ref{fig1}a):} The Agent processes a raw dataset of constitutive literature, encapsulating governing equations (e.g., thermal expansion, capillary laws) into discrete "Constitutive Skills" annotated with semantic metadata regarding interacting physical fields and applicability domains (e.g., saturation states).
    \item \textbf{Topological Construction (Figure \ref{fig1}b):} Synthesizing the user-specified simulation scenario (including geometry and boundary conditions like $T_{bnd}=200^\circ$C) with the skill library, the Agent autonomously assembles a causal graph. Crucially, it performs \textbf{Logical Pruning} to remove redundant mechanisms and \textbf{Intrinsic Completion} to introduce missing dissipation paths required by the boundary constraints.
    \item \textbf{Coupled Simulation (Figure \ref{fig1}c):} The constructed graph is compiled into a solver instance to execute the fully coupled simulation, with results verified against experimental stability criteria.
\end{enumerate}


\subsection{Autonomous Constitution via Emergent Causal Topology}
\begin{figure}[htbp]
    \centering

    \includegraphics[width=0.95\textwidth]{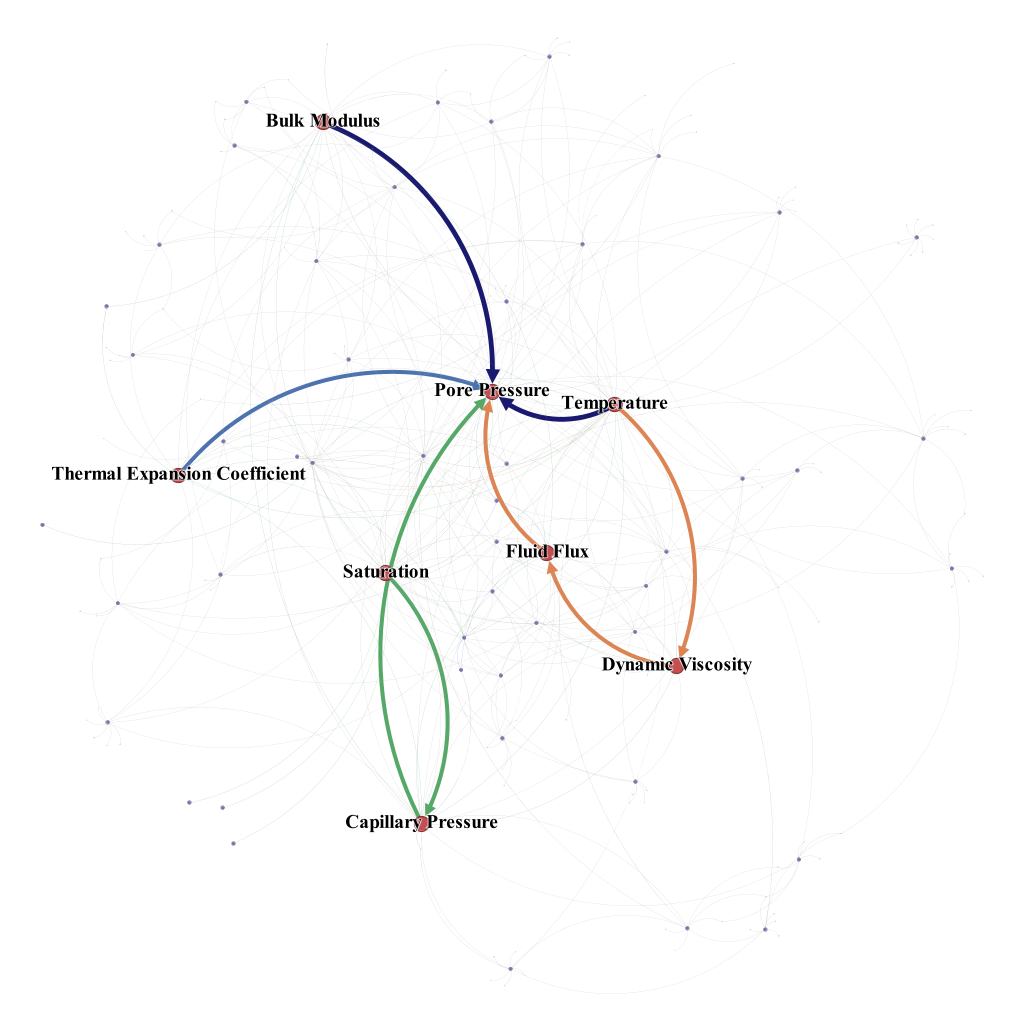}
    
    \caption{\textbf{Visualization of the implicit causal topology utilized by the Agent for autonomous mechanism construction.} 
    This graph is not an explicit input but an extracted representation of the Agent's underlying cognitive process, fusing \textbf{Intrinsic Priors} (LLM-based) with \textbf{Retrieved Skills} (literature-based). 
    \textbf{Dark Blue Path (Retrieved):} Represents the primary thermodynamic driver implicitly anchored to the Skill Library (\textit{Source II}), linking thermal expansion to pore pressure. 
    \textbf{Green Path (Pruned):} Represents the saturation-capillary constraint (\textit{Source I}). The Agent internally identifies this mechanism as valid but dynamically executes \textbf{pruning} due to the specific simulation context ($S_r=1$), filtering out mathematical redundancy. 
    \textbf{Orange Path (Intrinsic):} Represents a \textbf{mechanism completion} pathway derived from the LLM’s internal physical knowledge rather than external text. It reflects the Agent's latent capacity for fluid dissipation (Darcy flow), preserved to ensure physical stability against time-scale violations (discussed in Section \ref{sec2.3}).}
    
    \label{fig2}
\end{figure}

To transition from rigid, pre-defined solvers to an adaptive simulation paradigm, we introduce the concept of \textbf{"Constitutive Skills"}—encapsulated units of physical knowledge that bind mathematical governing equations with semantic metadata regarding their applicability and interacting fields. Figure \ref{fig2} visualizes the emergent causal topology constructed by the Agent, representing the dynamic assembly of these skills upon a substrate of intrinsic physical logic.

The framework operates through a context-aware construction process rather than static execution. Upon initialization with a specific simulation context, the Agent synthesizes a coupled system by orchestrating two distinct knowledge sources. First, it anchors critical driving forces to the Skill Library populated from domain literature. The Dark Blue Path (Figure \ref{fig2}), for instance, is not merely a connection but an instantiated operator, applying the thermal pressurization coefficient $\Lambda$ (extracted from Source II) to drive pore pressure evolution.

Crucially, this construction is governed by an autonomous optimization of physical fidelity and computational economy. The Agent evaluates the necessity of each candidate skill against the boundary conditions.

\begin{itemize}
    \item \textbf{Logical Pruning:} As illustrated by the \textbf{Green Path} (Figure \ref{fig2}), the Agent retrieves the saturation-capillary constraint (from Source I\cite{sun2018singularity}). However, by reasoning that the initial state is fully saturated ($S_r=1$), it identifies this mechanism as mathematically redundant and autonomously \textbf{prunes} the edge, simplifying the system without sacrificing accuracy.
    \item \textbf{Intrinsic Completion:} Conversely, the Agent recognizes that literature-derived skills may be insufficient. The Orange Path (linking Viscosity to Fluid Flux) emerges not from an external text, but from the LLM’s \textbf{intrinsic physical priors} (fluid mechanics fundamentals). The Agent identifies this pathway as a latent dissipation channel (Darcy flow), holding it in reserve to resolve potential physical inconsistencies that rigid literature assumptions (e.g., "undrained conditions") might overlook.
\end{itemize}

This dynamic topology serves as the blueprint for the subsequent automated coupling, ensuring that the final simulation is not just a stacking of equations, but a rationally constituted system tailored to the specific physical reality.

\subsection{Unit Verification of Constitutive Kernels}

The predictive fidelity of an autonomous simulation is fundamentally bounded by the accuracy of its constituent kernels. Before orchestrating the full multi-physics coupling, the Agent subjects the extracted literature-based skills—visualized as the Dark Blue and Green paths in Figure \ref{fig2}—to rigorous element-wise verification. This phase validates that the implicit causal links have been correctly instantiated into deterministic governing equations.

\textbf{Verification of High-Order Dynamics (The Green Path, Source I)}

\begin{figure}[htbp]
    \centering

    \includegraphics[width=0.95\textwidth]{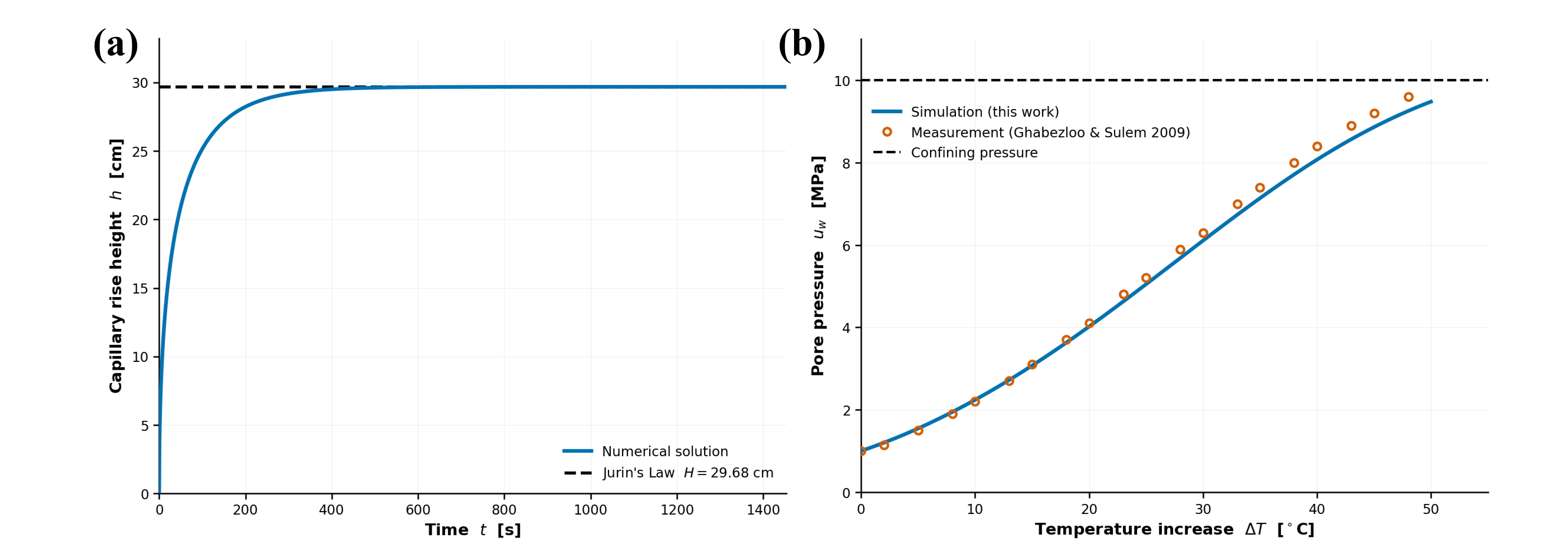} 
    
    \caption{\textbf{Element-wise verification of the literature-retrieved constitutive kernels.} 
    \textbf{a,} Validation of the capillary dynamics extracted from \textit{Source I} (The Green Path). The Agent-simulated capillary rise height $h(t)$ (solid blue line) accurately captures the transient rise and converges to the theoretical equilibrium limit defined by Jurin's Law (dashed black line, $H = 29.68$ cm). This confirms the Agent's capability to correctly instantiate the saturation-pressure relationship before it is autonomously pruned.
    \textbf{b,} Validation of the thermal pressurization driver extracted from \textit{Source II} (The Dark Blue Path). The simulation accurately reproduces the experimental pore pressure evolution from \textit{Ghabezloo \& Sulem (2009)} (orange markers) under undrained heating conditions. This establishes the accuracy of the primary thermodynamic driving force prior to the activation of the Agent's \textbf{intrinsic completion mechanism} (The Orange Path).}
    
    \label{fig3}
\end{figure}

Although the saturation-capillary mechanism is eventually pruned due to the specific initial condition ($S_r=1$), its verification serves as a baseline for the Agent's constitutive handling. As shown in Figure \ref{fig3}a, the Agent successfully reconstructs the capillary rise dynamics. The simulation captures the transient evolution and accurately converges to the equilibrium height ($H = 29.68$ cm) predicted by Jurin's Law. This confirms that the Agent correctly instantiates the physical laws governing fluid potential, validating the mathematical integrity of the Green Path prior to its context-aware pruning.

\textbf{Verification of the Thermodynamic Driver (The Dark Blue Path, Source II)}

The precision of the primary source term—thermal pressurization—is verified against experimental data from Ghabezloo \& Sulem (2009). Figure \ref{fig3}b illustrates the evolution of pore pressure ($u_w$) under undrained heating. The Agent’s simulation (solid line) exhibits a strong alignment with the experimental markers, confirming the accurate extraction and implementation of the thermal pressurization coefficient $\Lambda$. This result validates the integrity of the Dark Blue Path in the causal topology, ensuring that the external driving forces are physically sound before the Agent introduces its intrinsic completion mechanism to resolve system-level constraints.

\subsection{Context-Aware Reasoning via Dimensionless Scaling}\label{sec2.3}

Merely verifying constitutive kernels is insufficient, as the validity of literature-derived equations is often constrained by implicit assumptions—such as "undrained conditions"—that may not hold in the target simulation environment. To resolve this, the Agent employs dimensionless scaling analysis (specifically the Deborah number, $De$) as a cognitive filter to determine the dominant physical regime.

The Agent continuously evaluates the competition between the characteristic fluid diffusion time ($\tau_{diff}$) and the thermal loading timescale ($\tau_{load}$). This ratio is quantified as:

\begin{equation}
    De = \frac{\tau_{diff}}{\tau_{load}} \approx \frac{L^2 \mu_f(T) \beta}{k \cdot t}
\end{equation}

\textbf{Crucially, the Agent spontaneously isolates the Deborah number ($De$) as the unique governing parameter}, distinguishing it from other potential dimensionless candidates (e.g., Fourier number for heat conduction, Peclet number for advection). To explain this autonomous decision, our post-hoc analysis of the extracted knowledge graph (Figure \ref{fig2}) provides the mechanistic evidence. The graph reveals that the Agent internally structures pore pressure evolution as a competition between a Source Term (thermal pressurization rate, $\sim 1/\tau_{load}$) and a Diffusive Sink Term (Darcy flux, $\sim 1/\tau_{diff}$). By implicitly filtering out advection-dominated regimes ($Pe \to 0$) and intrinsic reaction kinetics ($Da$ not applicable), the Agent correctly deduces that the system's behavior is uniquely determined by the ratio of the intrinsic relaxation time ($\tau_{diff}$) to the extrinsic loading time ($\tau_{load}$). This ratio corresponds strictly to the Deborah number definition ($De = \tau_{diff}/\tau_{load}$), confirming that the Agent successfully derived the mathematically precise scaling law from its latent physical priors.

\begin{figure}[htbp]
    \centering
    \includegraphics[width=0.95\textwidth]{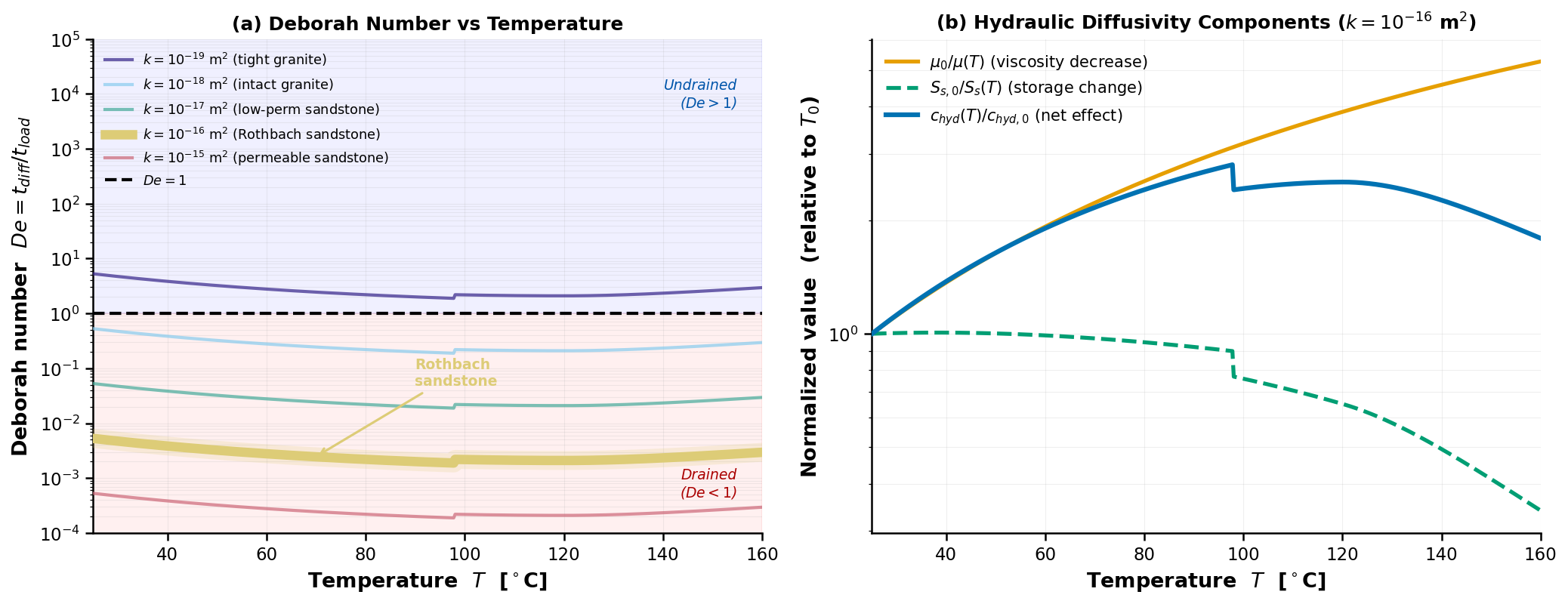} 
    
    \caption{\textbf{Autonomous regime detection via Deborah number scaling analysis.} 
    \textbf{a,} Evolution of the Deborah number ($De$) versus temperature for various rock permeabilities. The \textbf{bold yellow line} represents the target constitutive model used in this study (Rothbach sandstone, $k=10^{-16}$ m$^2$). The Agent identifies that the system operates entirely within the \textbf{Pink Zone ($De \ll 1$)}, signifying a "Drained" regime where fluid diffusion dominates. This scaling insight explicitly contradicts the "undrained" assumption often implied in simplified thermal pressurization models (Source II), compelling the Agent to activate intrinsic dissipation mechanisms.
    \textbf{b,} Decomposition of the hydraulic diffusivity components relative to their initial values at $T_0$. The Agent tracks the competing effects of viscosity reduction (Orange line, promoting flow) and storage capacity changes (Green line), synthesizing them into a net diffusivity trend (Blue line) to modulate the Darcy flow intensity dynamically.}
    \label{fig4}
\end{figure}

Figure \ref{fig4}a visualizes the Agent’s regime detection for the target material (Rothbach sandstone, $k=10^{-16}$ m$^2$, indicated by the \textbf{bold yellow line}).
Contrary to the "undrained" idealization ($De \gg 1$) often assumed in simplified theoretical treatments of thermal pressurization (Source II), the Agent identifies that the system actually operates deep within the \textbf{"Drained Regime"} (Pink Zone), where $De$ ranges from $10^{-2}$ to $10^{-3}$.
This scaling analysis reveals a critical physical reality: the timescale of fluid escape is two orders of magnitude faster than the heating rate.

\textbf{Mechanism Completion Trigger:}

This explicit identification of the $De \ll 1$ regime serves as the trigger for the Agent's \textbf{intrinsic mechanism completion}. The Agent recognizes that adhering strictly to an "undrained" formulation would lead to a gross overestimation of pore pressure. Consequently, it autonomously activates the Orange Path (Figure \ref{fig2})—the intrinsic Darcy flow mechanism—thereby injecting a diffusive dissipation term ($\nabla \cdot \mathbf{v}$) into the governing equations.
Furthermore, as shown in Figure \ref{fig4}b, the Agent accounts for the temperature-dependent evolution of hydraulic diffusivity components (viscosity $\mu_f$ and storage $S_s$), ensuring that the dissipation mechanism dynamically adapts to the thermal field. This process demonstrates the Agent's capacity to override literature-based simplifications with robust, scale-appropriate physical priors.

\subsection{Emergent Validity in Fully Coupled Simulations}
The culmination of the Agent's constitutive reasoning is the \textbf{fully coupled multi-physics simulation}, where the impact of its autonomous mechanism completion becomes undeniable. By integrating the verified thermal pressurization driver (Retrieved Skill) with the scale-appropriate fluid dissipation term (Intrinsic Prior), the Agent synthesizes a physically consistent model that prevents catastrophic prediction errors.

\begin{figure}[htbp]
    \centering
    \includegraphics[width=0.9\textwidth]{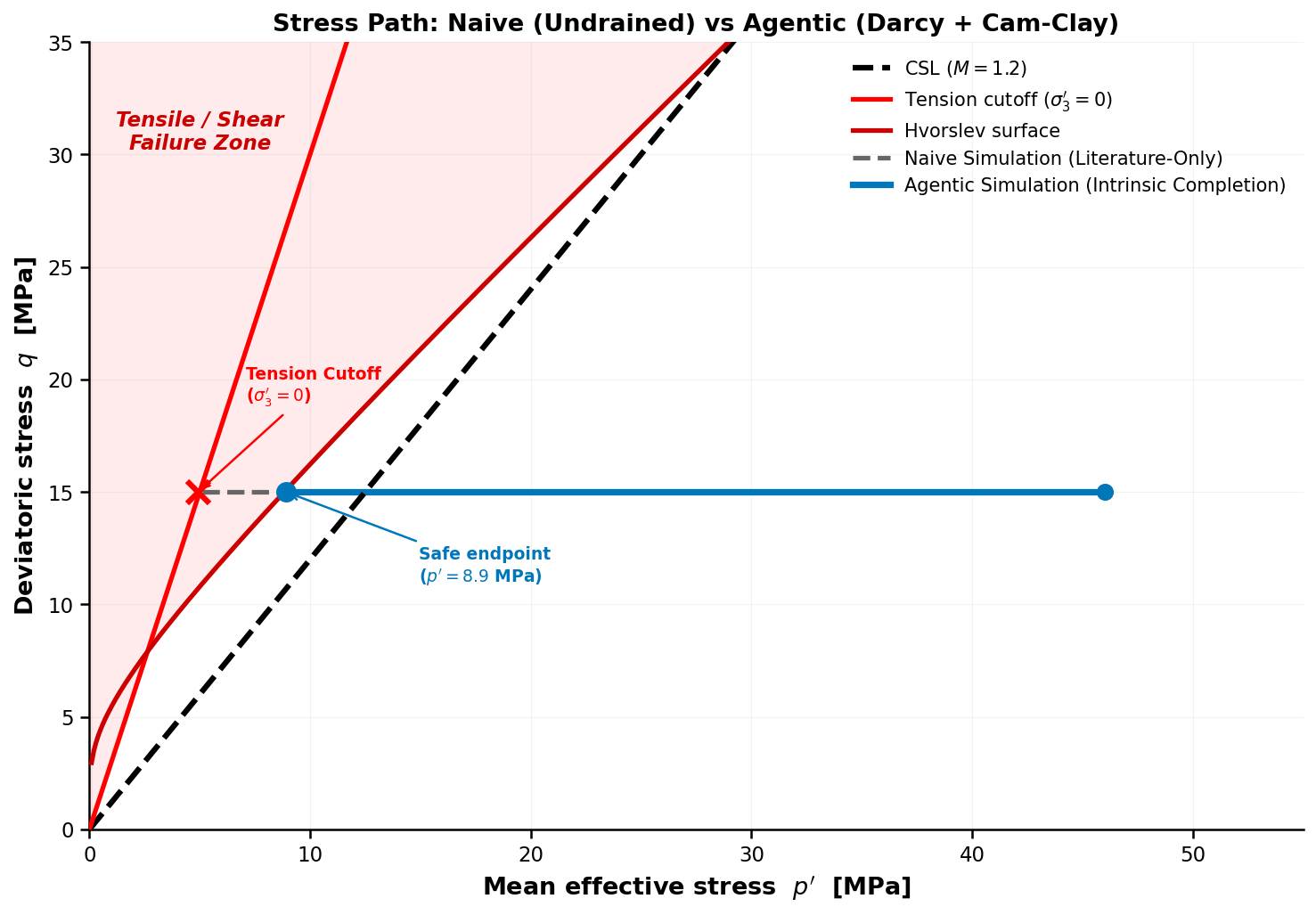} 
    
    \caption{\textbf{Emergent validity demonstrated via effective stress path stability.} 
    The plot tracks the evolution of the system in $p'-q$ space relative to the rock's failure envelopes (Red Zones: Tensile/Shear Failure).
    \textbf{Dashed Grey Line (Literature-Only):} Represents the model constrained strictly to the "undrained" literature extraction. The unchecked thermal pressurization drives the effective stress path into the failure zone (marked by the \textbf{Red Cross}), falsely predicting rock fracture.
    \textbf{Solid Blue Line (Agent-Completed):} Represents the fully coupled model where the Agent has autonomously activated its \textbf{intrinsic mechanism} (Darcy flow) upon detecting the $De \ll 1$ regime. The diffusive dissipation arrests the stress reduction, stabilizing the path at a \textbf{safe endpoint} ($p' = 8.9$ MPa) well within the elastic envelope.
    This comparison highlights that the Agent's mechanism completion is essential for preserving the \textbf{qualitative physical correctness} of the simulation.}
    \label{fig5}
\end{figure}

Figure \ref{fig5} visualizes this impact in the $p'-q$ stress space (mean effective stress vs. deviatoric stress), which serves as the definitive map for material stability.
The dashed grey line represents the naive "Literature-Only" simulation. Constrained to the "undrained" assumption extracted from Source II, the model predicts an unchecked accumulation of pore pressure. This forces the effective stress path to migrate rapidly to the left, eventually crossing the \textbf{Tension Cutoff} and \textbf{Hvorslev Surface} (Red Zone).
Critically, this naive model falsely predicts material failure, suggesting the rock would fracture under conditions where it is experimentally known to remain intact.

In stark contrast, the solid blue line represents the "Agent-Completed" simulation. Having diagnosed the Drained Regime ($De \ll 1$) via scaling analysis, the Agent autonomously activates the Darcy flow mechanism (The Orange Path).
The result is a stabilized stress path: the intrinsic dissipation relieves the excess pore pressure, arresting the leftward migration of effective stress. The system settles at a safe equilibrium endpoint ($p' = 8.9$ MPa), maintaining a sufficient safety margin from the failure envelopes.

This result serves as the definitive validation of "Emergent Validity." It confirms that without the Agent's autonomous reasoning to bridge the gap between literature assumptions and physical reality, the simulation would not merely be inaccurate—it would be \textbf{qualitatively wrong}, predicting structural failure where none exists.

\begin{figure}[htbp]
    \centering
    \includegraphics[width=0.9\textwidth]{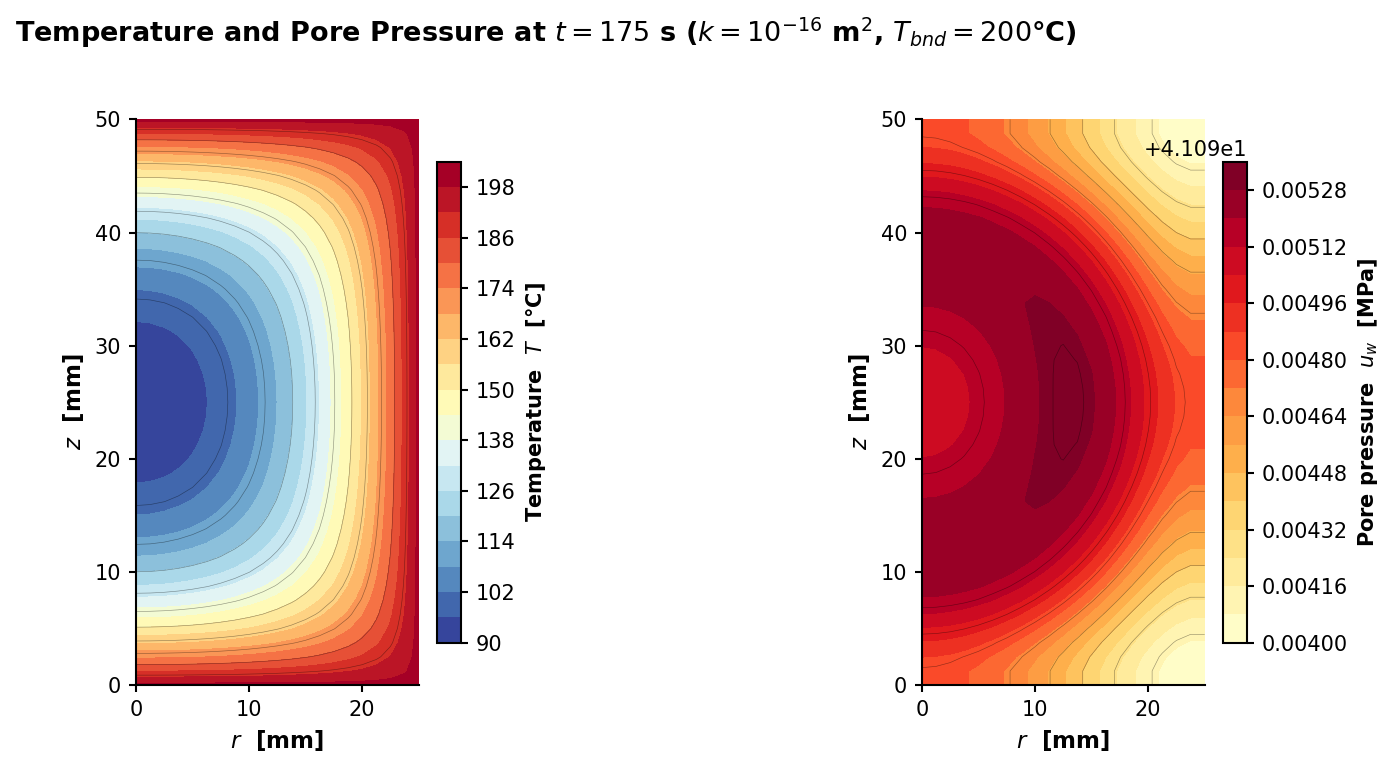} 
    
    \caption{\textbf{Spatial verification of the fully coupled solution via field contours.} 
    Snapshots of the system state at $t=175$ s for the target Rothbach sandstone ($k=10^{-16}$ m$^2$, $T_{bnd}=200^{\circ}$C).
    \textbf{a,} Temperature distribution ($T$), showing the conductive heating front propagating from the boundary ($r=25$ mm) towards the core.
    \textbf{b,} Pore pressure distribution ($u_w$). The pressure field is nearly uniform at $\approx 41.1$ MPa, with a subtle gradient driving fluid from the center (Dark Red) to the drainage boundary (Light Yellow). This confirms that the Agent's intrinsic mechanism ($\nabla \cdot \mathbf{v}$) effectively dissipates the thermally induced pressure, preventing the unphysical accumulation predicted by the naive model.}
    \label{fig6}
\end{figure}

Finally, the Agent's mechanism completion ensures the correct representation of spatial heterogeneity, a capability beyond the scope of zero-dimensional constitutive solvers. Figure \ref{fig6} presents the spatial fields of temperature ($T$) and pore pressure ($u_w$) at $t=175$ s for the Rothbach sandstone case ($k=10^{-16}$ m$^2$).
While the temperature field (Figure \ref{fig6}a) is governed by conductive heat transfer driving inwards from the boundary, the pore pressure field (Figure \ref{fig6}b) reveals the active role of the Agent-activated Darcy flow.
Crucially, the pore pressure stabilizes around 41.1 MPa (consistent with the stable effective stress $p' \approx 8.9$ MPa shown in Figure \ref{fig5}). The visible gradient—higher in the center and lower at the rim—confirms that the Agent successfully enforces the drainage boundary condition, allowing fluid to escape rather than accumulating to failure levels.

\section{Discussion}
\textbf{Epistemic Correction vs. Blind Translation}

The central contribution of this work is not merely the automation of code generation, but the demonstration of \textbf{epistemic correction}—the ability of an AI agent to challenge and rectify the assumptions embedded in its training data. Standard Large Language Models (LLMs) often function as "stochastic parrots," retrieving the most probable continuation of a text. Had our Agent acted as a standard Copilot, it would have strictly followed the retrieved literature (Source II), adopting the "undrained" assumption to produce code that runs without errors but fails physically (as shown by the "Naive" path in Figure \ref{fig5}). By explicitly employing Dimensionless Scaling Analysis (the Deborah number) as a cognitive gatekeeper, the Agent transcends linguistic translation. It exhibits a form of "physical intuition," recognizing that the mathematical validity of a closed-system equation breaks down under the open-system boundary conditions ($T_{bnd}$ active). This capability marks a paradigm shift from AI-as-Translator to AI-as-Physicist.

\textbf{The Role of Intrinsic Priors as Safety Valves}

The visualization of the latent knowledge graph (Figure \ref{fig2}) reveals why the Agent succeeds where rigid solvers fail. The "Intrinsic Priors"—specifically the connection between viscosity reduction and fluid flux (Darcy's Law)—act as a latent safety net. While the retrieved skills (RAG) provide the specific constitutive parameters for the material (Rothbach sandstone), the intrinsic priors provide the universal conservation laws required to maintain thermodynamic consistency. This Neuro-Symbolic interplay allows the Agent to "hallucinate" in a controlled, constructive manner: it invents a missing mechanism (the Orange Path) not because it was explicitly told to, but because the conservation of mass in a "Drained" regime ($De \ll 1$) demanded it. This suggests that future AI-for-Science frameworks should focus on curating these intrinsic conservation priors to prevent physical violations in generative simulation.

\textbf{Generalizability and Limitations}

While demonstrated on a specific Thermal-Hydro-Mechanical (THM) problem, the framework's logic—\textbf{Skill Encapsulation $\to$ Topological Pruning $\to$ Intrinsic Completion}—is agnostic to the specific physics. It can theoretically be generalized to other coupled domains, such as chemo-mechanics (battery degradation) or thermo-magnetism, provided the fundamental conservation laws are present in the LLM's latent space. However, the system's reliance on a pre-defined "Skill Library" (even if autonomously ingested) introduces a dependency on the quality of the seed literature. If the source texts contain fundamental errors in the functional form of equations (not just applicability assumptions), the current verification layer (Unit Testing) might not suffice. Future work will focus on integrating Data-Driven Discovery (e.g., symbolic regression on raw experimental data) to allow the Agent to correct not just the topology of the graph, but the kernel functions themselves.

\section{Methodology}
We introduce a Neuro-Symbolic Generative Framework capable of autonomously formulating and solving multi-physics problems where the governing laws are not predefined but dynamically synthesized from literature. Unlike standard "Copilots" that translate natural language instructions into rigid scripts, our system functions as a reasoning kernel, bridging the semantic gap between unstructured physical knowledge and rigorous numerical execution.
The workflow is orchestrated through three distinct phases: (1) Semantic Encapsulation, where raw equations are structured into functional units; (2) Physics-Informed Cognitive Reasoning, where the Large Language Model (LLM) implicitly filters and completes mechanisms based on dimensionless scaling analysis; and (3) Automated Variational Synthesis, where the validated physics is compiled into executable finite element code.

\subsection{Semantic Encapsulation of Constitutive Skills}
The foundational unit of the framework is the "Constitutive Skill," $\mathcal{S}$. Raw mathematical expressions extracted from literature (via RAG) are insufficient for autonomous reasoning as they lack context regarding their applicability and coupling requirements. To resolve this, we map each extracted law into a structured schema, transforming static equations into connectable semantic operators:
\begin{equation}
    \mathcal{S} = \{ \mathcal{F}, \Omega_{\text{int}}, \mathcal{C}_{\text{meta}} \}
\end{equation}
where:
\begin{itemize}
    \item $\mathcal{F}$ (Functional Form): The symbolic representation of the governing equation (e.g., $\epsilon_{th} = \alpha_T \Delta T$).
    \item $\Omega_{\text{int}}$ (Interaction Manifold): The set of physical fields required as inputs (prerequisites) and generated as outputs (targets), defining the node's connectivity potential.
    \item $\mathcal{C}_{\text{meta}}$ (Applicability Constraints): Semantic metadata defining the validity domain. This includes phase assumptions (e.g., "Saturated Liquid"), material symmetry, and thermodynamic restrictions.
\end{itemize}

This encapsulation allows the LLM to evaluate the utility of a skill not merely by keyword matching, but by logical compatibility with the simulation topology.

\subsection{Physics-Informed Cognitive Reasoning}
The core novelty lies in the Agent's ability to perform implicit topological reasoning. Upon receiving the user-specified simulation scenario (geometry, boundary conditions $\Gamma$, and initial state $\mathbf{u}_0$), the Agent does not simply stack all retrieved skills. Instead, it executes a \textbf{Chain-of-Thought (CoT)} reasoning process to synthesize a valid constitutive graph, employing two specific cognitive mechanisms:

\textbf{Deductive Logic: Constraint-Based Pruning}

The Agent first evaluates the compatibility between the retrieved skills' constraints $\mathcal{C}_{\text{meta}}$ and the scenario's initial conditions $\mathbf{u}_0$. Mechanisms identified as mathematically redundant or physically inapplicable are autonomously pruned.
Example: If a retrieved skill governs capillary pressure evolution ($P_c$) but the scenario specifies a fully saturated domain ($S_r = 1$), the Agent deduces the condition $\nabla P_c = 0$, thereby removing the corresponding term from the governing equations to prevent numerical singularities.

\textbf{Inductive Logic: Regime Identification via Scaling Analysis}

Beyond constraint satisfaction, the Agent validates the physical fidelity of the system using Dimensionless Scaling Analysis. Before formulation, the Agent estimates characteristic timescales (e.g., fluid diffusion $\tau_{\text{diff}}$ vs. thermal loading $\tau_{\text{load}}$) to compute key dimensionless numbers (such as the Deborah Number, $De$).
\begin{itemize}
    \item \textbf{Regime Detection:} The Agent classifies the dominant physical regime (e.g., $De \ll 1 \implies$ Drained; $De \gg 1 \implies$ Undrained).
    \item \textbf{Intrinsic Completion:} If the literature-derived equations imply an assumption (e.g., "Undrained") that contradicts the identified regime, the Agent detects a mechanism deficit. To restore physical consistency, it activates Intrinsic Priors—fundamental laws stored within the LLM's latent weights (e.g., Darcy's Law for dissipation)—and injects the missing divergence terms ($\nabla \cdot \mathbf{v}$) into the system.
\end{itemize}

\subsection{Automated Variational Synthesis}
Once the constitutive logic is finalized—pruned of redundancies and completed with intrinsic physics—the Agent proceeds to Solver Synthesis. This phase translates the abstract physical graph into a weak-form variational formulation executable by the FEniCSx finite element solver.

\begin{enumerate}
    \item Field Initialization: The Agent defines appropriate function spaces (e.g., Lagrange elements $P_2/P_1$) for the active fields identified in $\Omega_{\text{int}}$.
    \item Weak Form Derivation: The Agent symbolically converts the strong form of the coupled PDEs into the residual weak form:
$R(\mathbf{u}, \mathbf{v}) = \int_{\Omega} \left( \dots \right) d\Omega + \int_{\Gamma} \left( \dots \right) d\Gamma = 0$
where $\mathbf{v}$ represents the test functions. This derivation explicitly handles the integration of the intrinsic dissipation terms injected during the reasoning phase.
    \item Code Compilation: The finalized variational forms are compiled into a Python simulation script, encompassing mesh generation, time-stepping loops, and adaptive solver parameters tailored to the stiffness of the coupled system.
\end{enumerate}

\subsection{Visualization of Latent Physical Knowledge}

To validate that the Agent possesses the necessary physical priors for the reasoning described in Section 2.2, we developed a Knowledge Extraction Probe to visualize the implicit constitutive logic embedded within the LLM. It is important to note that the Agent's runtime workflow (Section 2.1–2.3) operates directly on latent representations; the explicit knowledge graph $G(V, E)$ presented in Figure \ref{fig2} is a post-hoc reconstruction designed to verify the model's internal consistency and coverage of multiphysics couplings.

We implemented a Three-Stage Probing Protocol to systematically externalize these latent dependencies:

\textbf{1. Seed-Based Breadth-First Expansion (Local Topology)}

Starting from a seed set of 11 fundamental state variables (e.g., Pore Pressure, Temperature), we query the LLM to identify the most critical immediate physical neighbors for each variable. This phase, analogous to a Breadth-First Search (BFS) with depth $d=2$, captures the primary constitutive laws (e.g., Terzaghi's principle linking $P$ to $\sigma'$).

\textbf{2. Combinatorial Pairwise Interrogation (Global Connectivity)}

To uncover indirect or long-range couplings that might be overlooked by local traversal, we generate all unique pairs from the seed set ($C_{11}^2 = 55$ pairs). The LLM acts as a discriminator for each pair, determining whether a physical pathway exists (either direct or mediated by an intermediate variable). This ensures that cross-disciplinary interactions (e.g., Temperature affecting Fluid Viscosity) are explicitly registered.

\textbf{3. Domain-Level Coupling Synthesis (Systemic Consistency)}

Finally, to verify the completeness of multi-physics interactions, we query the connectivity between entire physical sub-domains (T-H, H-M, T-M). This top-down pass ensures that macro-scale mechanisms, such as thermal pressurization, are correctly decomposed into valid edge attributes.

Raw outputs from the probing process are normalized via a Canonicalization Layer to map synonymous terms (e.g., "PWP" vs. "Pore Pressure") to a standardized ontology. The resulting graph (Figure \ref{fig2}) confirms that the LLM's internal knowledge structure inherently contains both the Driving Forces (e.g., Thermal Expansion) and the Dissipation Mechanisms (e.g., Darcy's Law) required for the autonomous reasoning demonstrated in the Results.

\subsection{Numerical Simulation Protocols}
To validate the emergent validity of the Agent's reasoning, we conducted comparative simulations of a cylindrical Rothbach sandstone specimen ($R=25$ mm, $H=50$ mm) subjected to rapid thermal loading. The study contrasts two distinct formulation strategies:

\begin{enumerate}
    \item Baseline "Naive" Model: A 0D representative elementary volume (REV) assuming strictly undrained conditions (derived from literature Source II\cite{ghabezloo2009stress}).
    \item Agent-Synthesized Model: A 2D axisymmetric finite difference model incorporating the intrinsic Darcy flow mechanism (derived from the Agent's reasoning).
\end{enumerate}

\textbf{Governing Constitutive Laws (Agent-Synthesized)}

\textbf{Mechanical Behavior (Modified Cam-Clay)}

The rock skeleton is modeled as an elasto-plastic continuum governed by the Modified Cam-Clay (MCC) yield surface in $p'-q$ space:
\begin{equation}
    f(p', q) = q^2 + M^2 p' (p' - p_c) = 0
\end{equation}
where $M=1.2$ is the critical state slope, and $p_c$ is the preconsolidation pressure evolving with plastic volumetric strain $\varepsilon_v^p$:
\begin{equation}
    \frac{dp_c}{p_c} = \frac{1+e_0}{\lambda - \kappa} d\varepsilon_v^p
\end{equation}
We incorporate a Tension Cutoff ($p' \ge 0$) and a Hvorslev Surface on the dry side ($p' < p_c/2$) to capture brittle failure modes typical of overconsolidated sandstone.

\textbf{Thermal-Hydraulic Coupling}

The pore pressure evolution is driven by the competition between thermal pressurization (source) and hydraulic diffusion (sink):
\begin{equation}
    \frac{\partial u_w}{\partial t} = \underbrace{\Lambda \frac{\partial T}{\partial t}}_{\text{Source}} + \underbrace{\nabla \cdot \left( \frac{k}{\mu_f(T) S_s} \nabla u_w \right)}_{\text{Sink}}
\end{equation}

\begin{itemize}
    \item \textbf{Source Term ($\Lambda$):} The thermal pressurization coefficient $\Lambda = (\alpha_f - \alpha_s)/(c_f + c_\phi)$ is dynamically calculated based on temperature-dependent fluid properties.
    \item \textbf{Sink Term (Darcy):} The hydraulic diffusivity $c_{hyd} = k/(\mu_f S_s)$ incorporates the temperature-dependent viscosity $\mu_f(T)$ following the Arrhenius law.
\end{itemize}

\textbf{Boundary and Initial Conditions}

The simulation initializes the specimen at geostatic equilibrium:
\begin{itemize}
    \item Initial State: $p'_0 = 46$ MPa, $q_0 = 15$ MPa, $u_{w,0} = 4.0$ MPa, $T_0 = 25^\circ$C.
    \item Mechanical BCs: Constant total stress ($\sigma_{radial}=45$ MPa, $\sigma_{axial}=60$ MPa) is maintained, mimicking a triaxial creep test.
    \item Thermal BCs: A Dirichlet boundary condition $T(t) = 25 + 1.0 \cdot t$ ($^\circ$C) is applied at the outer rim ($r=R$), inducing a radial thermal gradient.
    \item Hydraulic BCs: The Agent's model enforces a drainage boundary ($u_w = u_{w,0}$) at $r=R$, whereas the Naive model enforces a no-flux condition ($\nabla u_w \cdot \mathbf{n} = 0$) everywhere.
\end{itemize}

\textbf{Numerical Implementation via Operator Splitting}

The coupled system is solved using a staggered Operator Splitting scheme with a time step $\Delta t = 0.5$ s:

\begin{enumerate}
    \item Thermal Step: Solve the heat diffusion equation $\partial_t T = \alpha_{th} \nabla^2 T$ explicitly.
    \item Hydraulic Step: Update pore pressure $u_w$ by superimposing the local thermal pressurization increment $\Delta u_{TP} = B\Lambda \Delta T$ and solving the implicit Darcy diffusion equation.
    \item Mechanical Step: Update effective stress $\boldsymbol{\sigma}' = \boldsymbol{\sigma}_{total} - u_w \mathbf{I}$ and perform the closest-point projection return mapping algorithm to enforce the MCC plasticity and tension constraints.
\end{enumerate}

This rigorous numerical setup ensures that the divergence in results (Figure \ref{fig5}) arises solely from the presence or absence of the Agent-identified dissipation mechanism, isolating the cognitive contribution of the framework.

\begin{table}[htbp]
\centering
\caption{\textbf{Material parameters and simulation conditions for Rothbach sandstone.} Parameters are categorized by physical domain: Mechanical (M), Hydraulic (H), and Thermal (T). The asterisk (*) denotes temperature-dependent properties updated dynamically during the simulation.}
\label{tab:sim_params}
\begin{tabular}{l l c l}
\hline
\textbf{Parameter} & \textbf{Symbol} & \textbf{Value} & \textbf{Unit} \\
\hline
\multicolumn{4}{l}{\textit{Mechanical Properties (M)}} \\
Critical State Slope & $M$ & 1.2 & - \\
Preconsolidation Pressure & $p_{c0}$ & 60.0 & MPa \\
Normal Compression Index & $\lambda$ & 0.15 & - \\
Recompression Index & $\kappa$ & 0.03 & - \\
Initial Void Ratio & $e_0$ & 0.3 & - \\
Poisson's Ratio & $\nu$ & 0.3 & - \\
\hline
\multicolumn{4}{l}{\textit{Hydraulic Properties (H)}} \\
Intrinsic Permeability & $k$ & $1.0 \times 10^{-16}$ & m$^2$ \\
Fluid Viscosity* & $\mu_f(T)$ & Arrhenius & Pa$\cdot$s \\
Fluid Compressibility & $c_f$ & $4.0 \times 10^{-10}$ & Pa$^{-1}$ \\
Solid Compressibility & $c_s$ & $2.0 \times 10^{-11}$ & Pa$^{-1}$ \\
Specific Storage & $S_s$ & $1.28 \times 10^{-10}$ & Pa$^{-1}$ \\
\hline
\multicolumn{4}{l}{\textit{Thermal Properties (T)}} \\
Thermal Conductivity & $\lambda_T$ & 2.0 & W/(m$\cdot$K) \\
Specific Heat Capacity & $C_p$ & 1000 & J/(kg$\cdot$K) \\
Fluid Expansion Coeff.* & $\alpha_f(T)$ & $\sim 3.0 \times 10^{-4}$ & K$^{-1}$ \\
Solid Expansion Coeff. & $\alpha_s$ & $3.3 \times 10^{-5}$ & K$^{-1}$ \\
Thermal Diffusivity & $\alpha_{th}$ & $1.0 \times 10^{-6}$ & m$^2$/s \\
\hline
\multicolumn{4}{l}{\textit{Initial \& Boundary Conditions}} \\
Initial Confining Pressure & $\sigma_3'$ & 41.0 & MPa \\
Initial Pore Pressure & $u_{w,0}$ & 4.0 & MPa \\
Initial Temperature & $T_0$ & 25.0 & $^\circ$C \\
Heating Rate & $\dot{T}$ & 1.0 & $^\circ$C/s \\
\hline
\end{tabular}
\end{table}

\section{Conclusion}

We have introduced a Neuro-Symbolic Generative Agent capable of autonomously navigating the complexity of multi-physics simulation. By decoupling physical knowledge into semantic "Constitutive Skills" and orchestrating them through a Chain-of-Thought reasoning process, the Agent bridges the gap between unstructured scientific literature and rigorous numerical execution.

Three key conclusions emerge from this study:
\begin{enumerate}
    \item \textbf{Context-Aware Reasoning:} The Agent successfully identified that the "undrained" assumption prevalent in literature was invalid for the specific thermal loading rate applied to Rothbach sandstone, autonomously activating a Darcy dissipation mechanism to resolve the conflict.
    \item \textbf{Emergent Physical Validity:} The fully coupled simulation, synthesized without human intervention, predicted a safe effective stress path ($p' \approx 8.9$ MPa), preventing the false positive failure prediction ($p' \rightarrow 0$) characteristic of naive models.
    \item \textbf{Methodological Robustness:} The rigorous element-wise verification and regime detection mechanisms ensure that the AI's creativity is bounded by physical laws, effectively mitigating the risk of hallucination in scientific computing.
\end{enumerate}

This work demonstrates that AI can serve as more than a coding assistant; it can act as a reasoning partner, safeguarding simulations against the implicit cognitive biases and simplifications that often compromise manual modeling.

\section*{Data availability}
The datasets generated and/or analysed during the current study are available in the GitHub repository at \url{https://github.com/shuWuYue123/Neuro-Symbolic-Auto-Coupling}.

\section*{Code availability}
The source code for the Neuro-Symbolic Generative Agent and the simulation scripts are available in the GitHub repository at \url{https://github.com/shuWuYue123/Neuro-Symbolic-Auto-Coupling}.


\bibliographystyle{elsarticle-num}
\bibliography{references}

@article{sun2018singularity,
  title={Singularity-free approximate analytical solution of capillary rise dynamics},
  author={Sun, BoHua},
  journal={Science China. Physics, Mechanics \& Astronomy},
  volume={61},
  number={8},
  pages={084721},
  year={2018},
  publisher={Springer Nature BV}
}

@article{ghabezloo2009stress,
  title={Stress dependent thermal pressurization of a fluid-saturated rock},
  author={Ghabezloo, Siavash and Sulem, Jean},
  journal={Rock Mechanics and Rock Engineering},
  volume={42},
  number={1},
  pages={1--24},
  year={2009},
  publisher={Springer}
}

@article{jumper2021highly,
  title={Highly accurate protein structure prediction with AlphaFold},
  author={Jumper, John and Evans, Richard and Pritzel, Alexander and Green, Tim and Figurnov, Michael and Ronneberger, Olaf and Tunyasuvunakool, Kathryn and Bates, Russ and {\v{Z}}{\'\i}dek, Augustin and Potapenko, Anna and others},
  journal={nature},
  volume={596},
  number={7873},
  pages={583--589},
  year={2021},
  publisher={Nature Publishing Group UK London}
}

@article{merchant2023scaling,
  title={Scaling deep learning for materials discovery},
  author={Merchant, Amil and Batzner, Simon and Schoenholz, Samuel S and Aykol, Muratahan and Cheon, Gowoon and Cubuk, Ekin Dogus},
  journal={Nature},
  volume={624},
  number={7990},
  pages={80--85},
  year={2023},
  publisher={Nature Publishing Group UK London}
}

@article{raissi2019physics,
  title={Physics-informed neural networks: A deep learning framework for solving forward and inverse problems involving nonlinear partial differential equations},
  author={Raissi, Maziar and Perdikaris, Paris and Karniadakis, George E},
  journal={Journal of Computational physics},
  volume={378},
  pages={686--707},
  year={2019},
  publisher={Elsevier}
}

@inproceedings{li2021burigede,
  title={Burigede liu, Kaushik Bhattacharya, Andrew Stuart, and Anima Anandkumar. Fourier neural operator for parametric partial differential equations},
  author={Li, Zongyi and Kovachki, Nikola Borislavov and Azizzadenesheli, Kamyar},
  booktitle={International Conference on Learning Representations},
  volume={2},
  number={3},
  pages={4},
  year={2021}
}

@article{lu2021learning,
  title={Learning nonlinear operators via DeepONet based on the universal approximation theorem of operators},
  author={Lu, Lu and Jin, Pengzhan and Pang, Guofei and Zhang, Zhongqiang and Karniadakis, George Em},
  journal={Nature machine intelligence},
  volume={3},
  number={3},
  pages={218--229},
  year={2021},
  publisher={Nature Publishing Group UK London}
}

@article{wang2023scientific,
  title={Scientific discovery in the age of artificial intelligence},
  author={Wang, Hanchen and Fu, Tianfan and Du, Yuanqi and Gao, Wenhao and Huang, Kexin and Liu, Ziming and Chandak, Payal and Liu, Shengchao and Van Katwyk, Peter and Deac, Andreea and others},
  journal={Nature},
  volume={620},
  number={7972},
  pages={47--60},
  year={2023},
  publisher={Nature Publishing Group UK London}
}

@book{zienkiewicz2005finite,
  title={The Finite Element Method Set},
  author={Zienkiewicz, O.C. and Taylor, R.L.},
  isbn={9780080531670},
  year={2005},
  address={Oxford},
  publisher={Butterworth-Heinemann}
}

@book{versteeg2007introduction,
  title={An introduction to computational fluid dynamics the finite volume method, 2/E},
  author={Versteeg, Henk Kaarle},
  year={2007},
  publisher={Pearson Education India},
  address={New Delhi}
}

@article{boiko2023autonomous,
  title={Autonomous chemical research with large language models},
  author={Boiko, Daniil A and MacKnight, Robert and Kline, Ben and Gomes, Gabe},
  journal={Nature},
  volume={624},
  number={7992},
  pages={570--578},
  year={2023},
  publisher={Nature Publishing Group UK London}
}

@article{m2024augmenting,
  title={Augmenting large language models with chemistry tools},
  author={M. Bran, Andres and Cox, Sam and Schilter, Oliver and Baldassari, Carlo and White, Andrew D and Schwaller, Philippe},
  journal={Nature Machine Intelligence},
  volume={6},
  number={5},
  pages={525--535},
  year={2024},
  publisher={Nature Publishing Group UK London}
}

@article{wang2023voyager,
  title={Voyager: An open-ended embodied agent with large language models},
  author={Wang, Guanzhi and Xie, Yuqi and Jiang, Yunfan and Mandlekar, Ajay and Xiao, Chaowei and Zhu, Yuke and Fan, Linxi and Anandkumar, Anima},
  journal={arXiv preprint arXiv:2305.16291},
  year={2023}
}

@inproceedings{park2023generative,
  title={Generative agents: Interactive simulacra of human behavior},
  author={Park, Joon Sung and O'Brien, Joseph and Cai, Carrie Jun and Morris, Meredith Ringel and Liang, Percy and Bernstein, Michael S},
  booktitle={Proceedings of the 36th annual acm symposium on user interface software and technology},
  pages={1--22},
  year={2023}
}

@article{anthropic2024claude,
  title={The claude 3 model family: Opus, sonnet, haiku},
  author={Anthropic, AI},
  journal={Claude-3 Model Card},
  volume={1},
  number={1},
  pages={4},
  year={2024}
}

@article{team2023gemini,
  title={Gemini: a family of highly capable multimodal models},
  author={Team, Gemini and Anil, Rohan and Borgeaud, Sebastian and Alayrac, Jean-Baptiste and Yu, Jiahui and Soricut, Radu and Schalkwyk, Johan and Dai, Andrew M and Hauth, Anja and Millican, Katie and others},
  journal={arXiv preprint arXiv:2312.11805},
  year={2023}
}

@article{wei2022emergent,
  title={Emergent abilities of large language models},
  author={Wei, Jason and Tay, Yi and Bommasani, Rishi and Raffel, Colin and Zoph, Barret and Borgeaud, Sebastian and Yogatama, Dani and Bosma, Maarten and Zhou, Denny and Metzler, Donald and others},
  journal={arXiv preprint arXiv:2206.07682},
  year={2022}
}

@article{bubeck2023sparks,
  title={Sparks of artificial general intelligence: Early experiments with gpt-4},
  author={Bubeck, S{\'e}bastien and Chandrasekaran, Varun and Eldan, Ronen and Gehrke, Johannes and Horvitz, Eric and Kamar, Ece and Lee, Peter and Lee, Yin Tat and Li, Yuanzhi and Lundberg, Scott and others},
  journal={arXiv preprint arXiv:2303.12712},
  year={2023}
}

@article{wei2022chain,
  title={Chain-of-thought prompting elicits reasoning in large language models},
  author={Wei, Jason and Wang, Xuezhi and Schuurmans, Dale and Bosma, Maarten and Xia, Fei and Chi, Ed and Le, Quoc V and Zhou, Denny and others},
  journal={Advances in neural information processing systems},
  volume={35},
  pages={24824--24837},
  year={2022}
}

@article{yao2023tree,
  title={Tree of thoughts: Deliberate problem solving with large language models},
  author={Yao, Shunyu and Yu, Dian and Zhao, Jeffrey and Shafran, Izhak and Griffiths, Tom and Cao, Yuan and Narasimhan, Karthik},
  journal={Advances in neural information processing systems},
  volume={36},
  pages={11809--11822},
  year={2023}
}

@inproceedings{yao2022react,
  title={React: Synergizing reasoning and acting in language models},
  author={Yao, Shunyu and Zhao, Jeffrey and Yu, Dian and Du, Nan and Shafran, Izhak and Narasimhan, Karthik R and Cao, Yuan},
  booktitle={The eleventh international conference on learning representations},
  year={2022}
}

@article{schick2023toolformer,
  title={Toolformer: Language models can teach themselves to use tools},
  author={Schick, Timo and Dwivedi-Yu, Jane and Dess{\`\i}, Roberto and Raileanu, Roberta and Lomeli, Maria and Hambro, Eric and Zettlemoyer, Luke and Cancedda, Nicola and Scialom, Thomas},
  journal={Advances in Neural Information Processing Systems},
  volume={36},
  pages={68539--68551},
  year={2023}
}

@article{xi2025rise,
  title={The rise and potential of large language model based agents: A survey},
  author={Xi, Zhiheng and Chen, Wenxiang and Guo, Xin and He, Wei and Ding, Yiwen and Hong, Boyang and Zhang, Ming and Wang, Junzhe and Jin, Senjie and Zhou, Enyu and others},
  journal={Science China Information Sciences},
  volume={68},
  number={2},
  pages={121101},
  year={2025},
  publisher={Springer}
}

@article{shen2023hugginggpt,
  title={Hugginggpt: Solving ai tasks with chatgpt and its friends in hugging face},
  author={Shen, Yongliang and Song, Kaitao and Tan, Xu and Li, Dongsheng and Lu, Weiming and Zhuang, Yueting},
  journal={Advances in Neural Information Processing Systems},
  volume={36},
  pages={38154--38180},
  year={2023}
}

@inproceedings{brohan2023can,
  title={Do as i can, not as i say: Grounding language in robotic affordances},
  author={Brohan, Anthony and Chebotar, Yevgen and Finn, Chelsea and Hausman, Karol and Herzog, Alexander and Ho, Daniel and Ibarz, Julian and Irpan, Alex and Jang, Eric and Julian, Ryan and others},
  booktitle={Conference on robot learning},
  pages={287--318},
  year={2023},
  organization={PMLR}
}

@inproceedings{maeureka,
  title={Eureka: Human-Level Reward Design via Coding Large Language Models},
  author={Ma, Yecheng Jason and Liang, William and Wang, Guanzhi and Huang, De-An and Bastani, Osbert and Jayaraman, Dinesh and Zhu, Yuke and Fan, Linxi and Anandkumar, Anima},
  booktitle={The Twelfth International Conference on Learning Representations},
  year={2024}
}

@article{ji2023survey,
  title={Survey of hallucination in natural language generation},
  author={Ji, Ziwei and Lee, Nayeon and Frieske, Rita and Yu, Tiezheng and Su, Dan and Xu, Yan and Ishii, Etsuko and Bang, Ye Jin and Madotto, Andrea and Fung, Pascale},
  journal={ACM computing surveys},
  volume={55},
  number={12},
  pages={1--38},
  year={2023},
  publisher={ACM New York, NY}
}

@article{wang2023scibench,
  title={Scibench: Evaluating college-level scientific problem-solving abilities of large language models},
  author={Wang, Xiaoxuan and Hu, Ziniu and Lu, Pan and Zhu, Yanqiao and Zhang, Jieyu and Subramaniam, Satyen and Loomba, Arjun R and Zhang, Shichang and Sun, Yizhou and Wang, Wei},
  journal={arXiv preprint arXiv:2307.10635},
  year={2023}
}

@article{lewkowycz2022solving,
  title={Solving quantitative reasoning problems with language models},
  author={Lewkowycz, Aitor and Andreassen, Anders and Dohan, David and Dyer, Ethan and Michalewski, Henryk and Ramasesh, Vinay and Slone, Ambrose and Anil, Cem and Schlag, Imanol and Gutman-Solo, Theo and others},
  journal={Advances in neural information processing systems},
  volume={35},
  pages={3843--3857},
  year={2022}
}

@article{taylor2022galactica,
  title={Galactica: A large language model for science},
  author={Taylor, Ross and Kardas, Marcin and Cucurull, Guillem and Scialom, Thomas and Hartshorn, Anthony and Saravia, Elvis and Poulton, Andrew and Kerkez, Viktor and Stojnic, Robert},
  journal={arXiv preprint arXiv:2211.09085},
  year={2022}
}

@article{gaston2009moose,
  title={MOOSE: A parallel computational framework for coupled systems of nonlinear equations},
  author={Gaston, Derek and Newman, Chris and Hansen, Glen and Lebrun-Grandie, Damien},
  journal={Nuclear Engineering and Design},
  volume={239},
  number={10},
  pages={1768--1778},
  year={2009},
  publisher={Elsevier}
}

@misc{baratta_2025_18101307,
  author       = {Baratta, Igor A. and
                  Dean, Joseph P. and
                  Dokken, Jørgen S. and
                  Habera, Michal and
                  Hale, Jack S. and
                  Richardson, Chris N. and
                  Rognes, Marie E. and
                  Scroggs, Matthew W. and
                  Sime, Nathan and
                  Wells, Garth N.},
  title        = {DOLFINx: The next generation FEniCS problem
                   solving environment
                  },
  month        = dec,
  year         = 2025,
  publisher    = {Zenodo},
  doi          = {10.5281/zenodo.18101307},
}

@article{xue2023jax,
  title={JAX-FEM: A differentiable GPU-accelerated 3D finite element solver for automatic inverse design and mechanistic data science},
  author={Xue, Tianju and Liao, Shuheng and Gan, Zhengtao and Park, Chanwook and Xie, Xiaoyu and Liu, Wing Kam and Cao, Jian},
  journal={Computer Physics Communications},
  volume={291},
  pages={108802},
  year={2023},
  publisher={Elsevier}
}

@book{belytschko2014nonlinear,
  title={Nonlinear finite elements for continua and structures},
  author={Belytschko, Ted and Liu, Wing Kam and Moran, Brian and Elkhodary, Khalil},
  year={2014},
  publisher={John wiley \& sons},
  address={Chichester}
}

@article{sarker2022neuro,
  title={Neuro-symbolic artificial intelligence: Current trends},
  author={Sarker, Md Kamruzzaman and Zhou, Lu and Eberhart, Aaron and Hitzler, Pascal},
  journal={Ai Communications},
  volume={34},
  number={3},
  pages={197--209},
  year={2022},
  publisher={SAGE Publications Sage UK: London, England}
}

@article{udrescu2020ai,
  title={AI Feynman: A physics-inspired method for symbolic regression},
  author={Udrescu, Silviu-Marian and Tegmark, Max},
  journal={Science advances},
  volume={6},
  number={16},
  pages={eaay2631},
  year={2020},
  publisher={American Association for the Advancement of Science}
}

@article{krenn2022scientific,
  title={On scientific understanding with artificial intelligence},
  author={Krenn, Mario and Pollice, Robert and Guo, Si Yue and Aldeghi, Matteo and Cervera-Lierta, Alba and Friederich, Pascal and dos Passos Gomes, Gabriel and H{\"a}se, Florian and Jinich, Adrian and Nigam, AkshatKumar and others},
  journal={Nature Reviews Physics},
  volume={4},
  number={12},
  pages={761--769},
  year={2022},
  publisher={Nature Publishing Group UK London}
}

\end{document}